\documentclass[aip, jcp, amsmath, amsthm, amssymb, citeautoscript, reprint]{revtex4-2}
\usepackage{amsmath, amsthm, amssymb}
\usepackage{graphicx, float, xcolor, pgf}
\usepackage{physics, siunitx}
\usepackage[caption=false, justification=centerlast]{subfig}
\usepackage{gensymb}
\usepackage{hyperref}
\hypersetup{
    pdfauthor={Devansh Sharma and Amartya Bose},
    pdftitle={Path Integral Lindblad Dynamics in Presence of Time-Dependent Fields},
    colorlinks=true,
}
\usepackage[normalem]{ulem}

\allowdisplaybreaks
\begin{document}

\title{Path Integral Lindblad Dynamics in Presence of Time-Dependent Fields}
\author{Amartya Bose}
\email{amartya.bose@tifr.res.in}
\email{amartya.bose@gmail.com}
\affiliation{Department of Chemical Sciences, Tata Institute of Fundamental Research, Mumbai 400005, India}

\begin{abstract}
  The path integral Lindblad dynamics (PILD) method~[A. Bose, J. Phys. Chem. Lett. 15(12), 3363–3368 (2024)] had been introduced as a way of incorporating the impact of certain empirical processes like pumps and drains on the dynamics of quantum systems interacting with thermal environments. The method being based on the time-translational invariance of the Nakajima-Zwanzig memory kernel, however, was not able to account for time-dependent external fields. In this communication, we give an alternate, simpler formulation of PILD, that allows us to go beyond this limitation. It does not require the evaluation of the non-Markovian memory kernel directly, and consequently can be applied to Floquet systems as well.
\end{abstract}

\maketitle

Open quantum systems are characterized by a relatively
small-dimensional system of interest interacting, typically
energetically, with one or more thermal reservoirs. These reservoirs
are significantly larger in dimensionality.  Examples of problems that
may be modeled in this manner are ubiquitous ranging from charge to
exciton transport. Simulations of such systems have been made possible
by using methods that can evolve the reduce density
matrix. Approximate methods like the Lindblad master
equation~\cite{lindbladGeneratorsQuantumDynamical1976,
  goriniCompletelyPositiveDynamical1976} or the Bloch-Redfield master
equation~\cite{blochGeneralizedTheoryRelaxation1957} have been used in
different fields to achieve such simulations. Semiclassical
simulations provides another family of the feasible
algorithms~\cite{cottonSymmetricalQuasiClassicalSpinMapping2015}. However,
exact simulations are required for having systematically convergeable
results. Such simulations are based on the notions of Feynman-Vernon
influence functional~\cite{feynmanTheoryGeneralQuantum1963} the
hierarchical equations of motion
(HEOM)~\cite{tanimuraQuantumClassicalFokkerPlanck1991,
  tanimuraNumericallyExactApproach2020} or quasi-adiabatic propagator
path integral (QuAPI)
method~\cite{makriTensorPropagatorIterativeI1995,
  makriTensorPropagatorIterativeII1995}. Various developments over the
past years~\cite{strathearnEfficientNonMarkovianQuantum2018,
  makriSmallMatrixDisentanglement2020,
  boseMultisiteDecompositionTensor2022,
  bosePairwiseConnectedTensor2022,
  kunduEfficientMatrixFactorisation2020, xuTamingQuantumNoise2022}
have made simulations of large systems increasingly
feasible~\cite{kunduIntramolecularVibrationsExcitation2022,
  boseTensorNetworkPath2022}.

However, though this sort of a traditional setup allows for modulation
of dynamics through changes of system energy levels through the
system-environment interactions, it does not account for the various
extraction and pumping processes that could either decrease or
increase the number of quantum particles present in the system. These
features are present in several real-world transport systems and owing
to the coupled many-body nature of the dynamics, ignoring them leads
to qualitatively different dynamics. Computationally, it is often more
convenient to treat these processes empirically while treating the
more strongly interacting vibrational and solvent bath in a
numerically exact manner.

We discuss two different examples, which provide slightly different
motivations to the empirical treatment. As a first example, consider
one of the many light-harvesting complexes which constitute the first
part of the photosynthetic process. These so-called antenna complexes
shuttle the excitation generated at one end of the complex to the sink
point where the excitation is dumped onto the special pair of
chlorophylls, and subsequently taken into the reaction center where
the exciton gets split into an electron hole pair and further
reactions happens. The entry of the exciton into the special pair is
practically a one-way transfer. To truly describe all of this
dynamics, one needs to include all the chromophores in the special
pair and then the sites in the reaction center. However, that leads to
an exponential growth of computational costs. So long as we are only
interested in the transport process in the antenna complex, it is
convenient to think of this as an empirical loss process from the
antenna complex.

The second type of process is where there is a non-innocent
environment which actively participates in the transport process
instead of just changing the energy of the states. Consider a
excitonic polariton where the cavity is lossy. The photon can leak out
of the cavity decaying into the continuum of environmental
electromagnetic modes. This is also difficult to characterized, and we
would like to simplify the description as an empirical loss.

Empirical treatments, therefore, need to be incorporated in the exact
dynamics methods. One possible route is to describe loss processes
through non-Hermitian descriptions of systems in a path integral
simulation~\cite{palmNonperturbativeEnvironmentalInfluence2017,
  sharmaNonHermitianStatetoStateAnalysis2025}. However, presence of
non-Hermiticity makes the time-propagation non-unitary, leading to
lack of trace-preservation. There is a definite requirement of methods
that maintain the completely positive, trace preserving (CPTP) aspect
of time-evolution of the reduced density matrix. This can be achieved
by incorporating the empirical processes through Lindblad jump
operators while continuing to provide a numerically exact description
of the thermal environment using path integrals and Feynman-Vernon
influence functional\cite{feynmanTheoryGeneralQuantum1963}. The path
integral Lindblad dynamics (PILD)
method\cite{boseIncorporationEmpiricalGain2024} allows for exactly
this combination to be achieved. An added benefit of the CPTP
propagation ensured by PILD is a more correct description of spectra
in presence of loss
mechanisms~\cite{sharmaImpactLossMechanisms2024}. Moreover, PILD can
account for a more general class of empirical processes including but
not limited to pumping mechanisms which cannot be described using
non-Hermitian methods. However, as formulated, the algorithm of
PILD~\cite{boseIncorporationEmpiricalGain2024} is currently unable to
handle systems interacting with time-dependent fields. In this
communication, we lift this restriction, and demonstrate a
implementation of PILD that can both simplify simulations and allow
for incorporation of time-dependence in the system Hamiltonian.

Consider a system described by the Hamiltonian, $\hat{H}_0$,
interacting with thermal baths
\begin{align}
  \hat{H} &= \hat{H}_0 + \sum_b\sum_j \frac{p_{bj}^2}{2} + \frac{1}{2}\omega_{bj}^2\left(x_{bj} - \frac{c_{bj}\hat{s}_b}{\omega_{bj}^2}\right)^2
\end{align}
where the frequencies and couplings, $\omega_{bj}$ and $c_{bj}$, of
the $j$th mode of the $b$th bath are specified by a spectral density
$J_b(\omega) = \frac{\pi}{2}\sum_j
\frac{c_{bj}^2}{\omega_{bj}}\delta(\omega-\omega_{bj})$. Additionally
this system is also subject to the influence of certain processes
described empirically by Lindblad jump operators, $L_n$. PILD
simulates the time-evolution of the reduced density matrix of this
system by solving the Nakajima-Zwanzig master
equation~\cite{nakajimaQuantumTheoryTransport1958,
  zwanzigEnsembleMethodTheory1960} augmented to take into account the
Lindblad jump operators~\cite{lindbladGeneratorsQuantumDynamical1976}:
\begin{align}
    \dv{\tilde\rho}{t} &= -\frac{i}{\hbar}\mathcal{L}_0\tilde\rho(t) + \int_0^{\tau_\text{mem}} \mathcal{K}(\tau)\tilde\rho(t-\tau)\dd{\tau}\nonumber\\
    &+ \sum_n L_n\tilde\rho(t)L_n^\dag - \frac{1}{2}\acomm{L_n^\dag L_n}{ \tilde\rho(t)}\label{eq:pild}
\end{align}
where $\mathcal{L}_0 = \comm{\hat{H}_0}{\cdot}$ is the bare system
Liouvillian operator and $\mathcal{K}$ is the memory kernel
corresponding to the thermal environments. Recently a lot of effort
has gone into estimating the memory kernel through semiclassical
simulations~\cite{mulvihillCombiningMappingHamiltonian2019,
  mulvihillRoadMapVarious2021}, but it is a non-trivial process.

PILD, as developed~\cite{boseIncorporationEmpiricalGain2024}, uses any
simulation method, say
QuAPI~\cite{makriTensorPropagatorIterativeI1995} or the time-evolving
matrix product operators
(TEMPO)~\cite{strathearnEfficientNonMarkovianQuantum2018}, to generate
the dynamical map, $\mathcal{E}(t)$, which can be used to propagate
the system's reduced density matrix
$\tilde\rho(t)=\mathcal{E}(t)\tilde\rho(0)$ in absence of the Lindblad
jump operators. Using this time-series of dynamical maps,
$\mathcal{E}(t)$, the transfer
tensors~\cite{cerrilloNonMarkovianDynamicalMaps2014} that satisfy
\begin{align}
  \tilde\rho(t) = \sum_{j=1}^L T_j\tilde\rho(t - j\Delta t)
\end{align}
are computed. These transfer tensors can be thought of as
time-discretized versions of the memory kernel:
\begin{align}
  T_k &= (1 - i\hat{\mathcal{L}}_0\Delta t)\delta_{k,1} + \mathcal{K}_k\Delta t^2.\label{eq:TTM-to-memkernel}
\end{align}

However, both the mapping of the dynamical maps onto the memory
kernels, Eq.~\eqref{eq:TTM-to-memkernel}, and the form of the
augmented Nakajima-Zwanzig master equation in Eq.~\eqref{eq:pild}
assumes a time-independent system Hamiltonian and a separable initial
condition. A different formulation that bypasses the Nakajima-Zwanzig
master equation is required to be able to incorporate the effect of
time-dependent fields.

Using Feynman-Vernon influence
functional~\cite{feynmanTheoryGeneralQuantum1963}, the time evolution
of the reduced density matrix in absence of the Lindblad jump
operators is represented by~\cite{makriTensorPropagatorIterativeI1995,
  makriTensorPropagatorIterativeII1995}:
\begin{align}
  \mel{s_N^+}{\tilde\rho(N\Delta t)}{s_N^-} &= \sum_{\left\{s_j^\pm\right\}}\mel{s_N^+}{\hat{U}}{s_{N-1}^+}\mel{s_{N-1}^+}{\hat{U}}{s_{N-2}^+}\nonumber\\
                                            &\times\cdots\mel{s_1^+}{\hat{U}}{s_0^+}\mel{s_0^+}{\tilde\rho_0}{s_0^-}\nonumber\\
                                            &\times\mel{s_0^-}{\hat{U}^\dag}{s_1^-}\cdots\mel{s_{N-1}^-}{\hat{U}^\dag}{s_N^-}\nonumber\\
  &\times F\left[\left\{s_j^\pm\right\}\right]\label{eq:pi}
\end{align}
where $\hat{U}=\exp(-i\hat{H}_0\Delta t/\hbar)$ is the short-time
propagator for the bare system. The forward and the backward
short-time propagator matrix elements are often considered together as
one unit coming from a super-operator
$\mel{s_{j+1}^\pm}{\hat{K}}{s_j^\pm} =
\mel{s_{j+1}^+}{\hat{U}}{s_j^+}\times\mel{s_j^-}{\hat{U}^\dag}{s_{j+1}^-}
= \mel{s^\pm_{j+1}}{\exp(-\frac{i}{\hbar}\hat{\mathcal{L}}_0\Delta
  t)}{s^\pm_j}$. The Feynman-Vernon influence functional, which
captures the interaction of the system and the environment in a
non-perturbative manner, is $F$. This term makes the simulations
non-Markovian.

To incorporate the jump operators, one has to revisit the definition
of the bare short-time propagator and think in terms of the
forward-backward propagator super-operator or the bare dynamical
map. In this case, in absence of the environment, the density matrix
of the system should evolve under the Lindblad master equation,
\begin{align}
    \dv{\tilde\rho}{t} &= -\frac{i}{\hbar}\hat{\mathcal{L}}_0\tilde\rho(t)  + \sum_n L_n\tilde\rho(t)L_n^\dag - \frac{1}{2}\acomm{L_n^\dag L_n}{ \tilde\rho(t)},\label{eq:lindblad}
\end{align}
which can be formally solved in the Liouville space as
$\tilde\rho(t+\Delta t) = K(\Delta t) \tilde\rho(t)$. The
super-operator $\hat{K} = \exp(-i\hat{\mathcal{L}}\Delta t/\hbar)$ is
related to the total Liouvillian,
$\hat{\mathcal{L}} = \hat{\mathcal{L}}_0 +
\hat{\mathcal{L}}_\text{Lindbladian}$, where
$\hat{\mathcal{L}}_0$ is the part of the Liouvillian
generated by the system Hamiltonian, and
$\hat{\mathcal{L}}_\text{Lindbladian}$ is the part that is generated
by the jump operators. Assuming a row-major vectorization of the
density matrix, we can easily show that these terms can be expressed
as
\begin{align}
    \hat{\mathcal{L}}_0 &= \hat{H}_0\otimes\mathbb{I} - \mathbb{I}\otimes\hat{H}^*_0\\
    \hat{\mathcal{L}}_\text{Lindbladian} &= i\hbar \sum_n L_n\otimes L_n^*\nonumber\\
    &- \frac{i\hbar}{2}\sum_n\left(L_n^\dag L_n \otimes\mathbb{I} + \mathbb{I}\otimes L_n^T L_n^*\right).
\end{align}
Here $\mathbb{I}$ is the identity operator of dimensions commensurate
with the system dimensionality. (In the row-major representation of
the density matrix, if an operator, $A$, acts on the density matrix
from the left, it gets mapped to the super-operator
$A\otimes\mathbb{I}$. Similarly, if an operator, $B$, acts from the
right, it gets mapped to the super-operator $\mathbb{I}\otimes B^T$.)
Once the Lindbladian is obtained, the bare forward-backward
propagator, $\hat{K}$, can be directly used in Eq.~\eqref{eq:pi}
instead of the forward and backward propagators $\hat{U}$ and
$\hat{U}^\dag$.

This can be trivially extended to the case of time-dependent
Hamiltonians or systems interacting with time-dependent fields by
solving the equation of motion,
\begin{align}
    \dv{\hat{K}}{t} &= -\frac{i}{\hbar}\hat{\mathcal{L}}(t)\hat{K}(t),
\end{align}
where
$\hat{\mathcal{L}}=\hat{\mathcal{L}}_0(t) +
\hat{\mathcal{L}}_\text{Lindbladian}$, solved with the initial
condition $\hat{K}(0) = \mathbb{I}$ in the Liouville space. In fact,
the time-dependence could in principle also be in the Lindblad jump
operators. If the time-constants or ``decay rates'' corresponding to
the jump operators are negative, they can encode non-Markovianity in
the empirical processes as
well\cite{ivanchenkoInvestigatingNonMarkovianEffects2025}.

\begin{figure}
  \centering
  \includegraphics{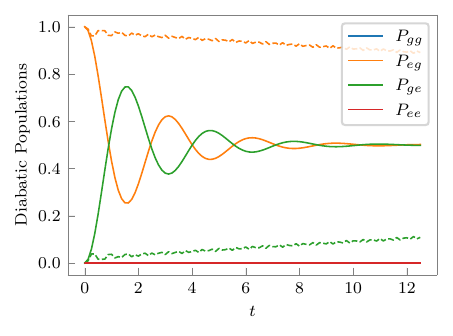}
  \caption{Population dynamics of various states without Lindblad jump operators. Solid lines: no external field; Dashed line: with external field.}\label{fig:no-lindblad}
\end{figure}

As an illustration consider a model excitonic dimer described by the
Hamiltonian:
\begin{align}
  \hat{H}_0 &= \epsilon_1\dyad{eg} + \epsilon_2\dyad{ge} + (\epsilon_1 + \epsilon_2)\dyad{ee} - \Delta \dyad{eg}{ge}
\end{align}
with $\epsilon_1 = \epsilon_2 = 5$ and $\Delta = 1$. The vibrations on
each of the monomers are modeled by two identical Ohmic baths with
$J(\omega) = \frac{\pi}{2}\hbar\xi\omega\exp(-\omega/\omega_c)$ with
$\xi=0.16$ and $\omega_c=7.5$. We compare the dynamics with and
without an external field
$V(t) = 11.96575 \cos(10t) \left(\dyad{eg} - \dyad{ge}\right)$. Before
exploring the effect of the empirical processes on this system, we
demonstrate the dynamics starting from $\tilde\rho(0) = \dyad{eg}$ in
Fig.~\ref{fig:no-lindblad}. The dynamics is localized in the first
excitation subspace because of the block diagonal structure of the
Hamiltonian. In persence of the external field, we see a localization
of the population on the $\dyad{eg}$ state in the resultant Floquet
system.

\begin{figure}
  \centering
  \subfloat[Population of diabatic states]{\includegraphics{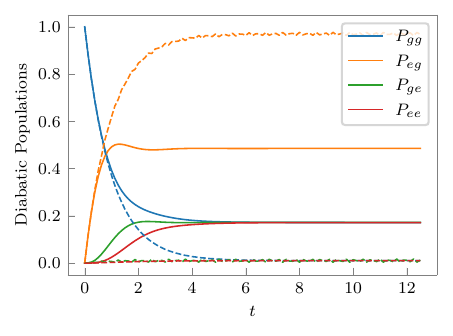}}

  \subfloat[Population of monomers]{\includegraphics{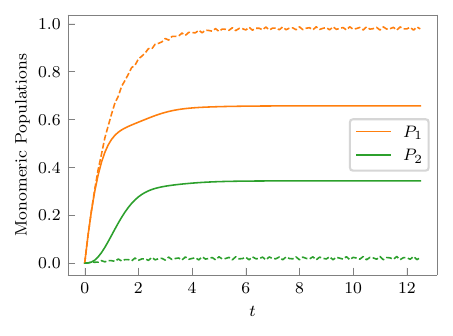}}
  \caption{Population dynamics of various states in presence of simultaneous pumping and draining. Solid lines: no external field; Dashed line: with external field.}\label{fig:lindblad}
\end{figure}

Now let us study the same system under conditions where the left
monomer is pumped and the right monomer is drained by the Lindbladian
jump operators
\begin{align}
  L_1 &= \dyad{eg}{gg} + \dyad{ee}{ge}\\
  L_2 &= \dyad{gg}{ge} + \dyad{eg}{ee}.
\end{align}
The dynamics starting from $\tilde\rho(0) = \dyad{gg}$ is shown in
Fig.~\ref{fig:lindblad} both in the diabatic basis and in the
monomeric basis where
$P_1(t) = \mel{eg}{\tilde\rho(t)}{eg}+\mel{ee}{\tilde\rho(t)}{ee}$ and
$P_2(t) = \mel{ge}{\tilde\rho(t)}{ge} +
\mel{ee}{\tilde\rho(t)}{ee}$. Notice that in absence of the
time-dependent field, though we are starting with no excitation in the
system, the Lindblad pumping mechanism starts to populate the
$\ket{eg}$ state. As the excitation hops from $\ket{eg}$ to $\ket{ge}$
over time, and eventually leaks out from there, the first monomer
again gets excited, potentially taking the system to the doubly
excited $\ket{ee}$ state. The net result of all this is the
establishment of a steady state.

In absence of the pumping and draining Lindblads, we saw that the
time-dependent field $V(t)$ has a tendency of localizing the
excitation on the left monomer. The same feature happens in the
presence of the Lindblads as well. It effectively slows down the
transfer of the excitation from the first to the second monomer. A
combination of this stalling of transport into the second monomer and
the leakage from there means that the net excited state population on
the second monomer is basically zero.

In this communication, we have provided an alternate formulation of
the PILD method\cite{boseIncorporationEmpiricalGain2024} that is not
dependent on the evaluation of the non-Markovian memory kernel,
allowing for the simultaneous incorporation of empirical processes
along with the impact of time-dependent fields on the system. The
present work along with the recently derived extension of the
state-to-state analysis\cite{boseImpactSolventStatetoState2023} to
PILD\cite{sharmaRoutesTransportPath2025} would allow for mechanistic
explorations of even more complex open floquet systems. The code
implementing PILD will be published later as a part of the
QuantumDynamics.jl
package\cite{boseQuantumDynamicsjlModularApproach2023}.

\bibliography{library}
\end{document}